\documentclass[twocolumn]{article}
\usepackage{graphicx}
\usepackage{titlesec}
\usepackage{amsmath,amssymb,bm,subfigure,color,url,booktabs,mathrsfs,cite}
\usepackage[colorlinks=true,citecolor=amazonite,linkcolor=amazonite,urlcolor=amazonite]{hyperref}

\definecolor{amazonite}{RGB}{0,115,150}
\definecolor{myred}{RGB}{255,56,0}
\definecolor{mygreen}{RGB}{30,150,30}
\definecolor{mybrown}{RGB}{150,30,30}
\definecolor{darkblue}{RGB}{30,50,100}
\definecolor{darkred}{RGB}{80,30,30}

\titleformat*{\section}{\centering\bfseries}
\titleformat*{\subsection}{\bfseries}
\titleformat*{\subsubsection}{\it}
\titleformat*{\paragraph}{\it}

\AtBeginDocument{
\textwidth=18.51cm
\textheight=23.5cm
\oddsidemargin=-1cm
\evensidemargin=-1cm
\headheight=0cm
\headsep=0cm
\topmargin=-0.7cm
\footskip=1cm
\columnsep=0.7cm

\heavyrulewidth=.08em
\lightrulewidth=.05em
\cmidrulewidth=.03em
\belowrulesep=.65ex
\belowbottomsep=0pt
\aboverulesep=.4ex
\abovetopsep=0pt
\cmidrulesep=\doublerulesep
\cmidrulekern=.5em
\defaultaddspace=.5em
}

\begin{document}

\twocolumn[{
\centering
{\large\bf
Dynamic cluster structure and predictive modelling of music creation style distributions
}

\vskip 0.4cm
Rajsuryan Singh$^1$ and Eita Nakamura$^{2,3}$
\vskip 0.2cm

{\it\small
$^1$Music Technology Group, Universitat Pompeu Fabra, 08002 Barcelona, Spain\\ 
$^2$The Hakubi Center for Advanced Research, Kyoto University, Kyoto 606-8501, Japan\\
$^3$Graduate School of Informatics, Kyoto University, Kyoto 606-8501, Japan\\
}
\vskip 0.7cm
}]

\begin{abstract}
We investigate the dynamics of music creation style distributions to understand cultural evolution involving advanced intelligence. Using statistical modelling methods and several musical statistics extracted from datasets of popular music created in Japan and the United States (the US), we explored the dynamics of cluster structures and constructed a fitness-based evolutionary model to analyze and predict the evolution of music creation style distributions. We found that intra-cluster dynamics, such as the contraction of a cluster and the shift of a cluster centre, as well as inter-cluster dynamics represented by clusters' relative frequencies, often exhibit notable dynamical modes that hold across the cultures and different musical aspects. Additionally, we found that the evolutionary model incorporating these dynamical modes is effective for predicting the future creation style distributions, with the predictions of cluster frequencies and cluster variances often having comparable contributions. Our results highlight the relevance of intra-cluster dynamics in cultural evolution, which have often been overlooked in previous studies.
\end{abstract}

\vspace{-2mm}
\paragraph{Keywords:}
Cultural evolution; statistical learning; music creation style; dynamical system; social dynamics; trend prediction.
\vspace{-2mm}

\section{Introduction}
\label{sec:Intro}

The theory of cultural evolution uses mathematical models to understand how knowledge and intelligent behaviours are shared and developed in human society \cite{CavalliSforzaFeldman,BoydRicherson}.
To construct models of cultural evolution, it is necessary to understand the processes by which knowledge or behaviour is transmitted and modified over generations \cite{Baxter2006,Griffiths2007,Smith2008,Nakamura2021}.
Some creative cultures, such as science and the arts involving advanced intelligence, are developed through the ability to learn to create complex artefacts.
Therefore, a quantitative understanding of the dynamic changes in the knowledge for creating artefacts is essential to uncover the mechanisms of the evolution of creative cultures.
Owing to its significance and universal presence in the society \cite{Mehr2019}, music has been a popular topic in cultural evolution research \cite{Serra2012,Zivic2013,Mauch2015,Park2015,Weiss2018,Interiano2018,Nakamura2019,Youngblood2019,Moss2022,Savage2022}.
Here, we study the evolution of music styles using large-scale data analysis and computational modelling.

Music composition has long been studied from the perspective of information processing.
The currently accepted view is that the knowledge necessary for music composition is acquired through learning the statistics extracted from musical products \cite{Fernandez2013}.
For example, a number of automatic music composition systems capable of generating songs similar to human-made ones have been realized based on statistical learning \cite{Pachet2011,Fukayama2010}.
Additionally, cognitive scientific experiments have also indicated that humans employ statistical learning to acquire knowledge of music and language \cite{Saffran1996,Ettlinger2011}.
Moreover, statistical learning has been used as a model to explain how humans learn to understand music in tasks such as music expectation \cite{Pearce2006}, genre classification \cite{PerezSancho2009}, and musical analysis \cite{Mavromatis2012}.
From this perspective, music creation is described as a process of generating sequential data using a probabilistic model, with the knowledge used in this process represented by the statistical parameters of the model.
For example, in the Markov model, a standard model for melody generation \cite{Ames1987}, the parameters represent the transition probabilities of pitches or other musical elements.
This formalization enables quantitative comparisons of creation styles between songs or composers, and accordingly allows a mathematical description of the social-level dynamics of the creative culture in terms of the time evolution of the distribution of the statistical parameters.
In the following, we identify these statistical parameters as quantified creation styles.

Musicologists have studied how musical features have changed over time.
Traditionally, researchers have provided detailed accounts of composition techniques within the social context \cite{HistoryOfMusicalStyle,HistoryOfWesternMusic,Negus1996,Holt2007}; however, most analyses lacked quantitative formulations that enable prediction and hypothesis testing.
Owing to the increasing availability of digital music data and computational analysis methods, large-scale quantitative analyses have recently become possible.
Previous studies have confirmed that in the history of Western classical and popular music, the means of various musical features evolved steadily and directionally \cite{Serra2012,Zivic2013,Mauch2015,Weiss2018,Interiano2018,Nakamura2019}.
Further, these evolutions were punctuated by periods of rapid change (i.e.\ revolutions) \cite{Zivic2013,Mauch2015,Weiss2018}, justifying the widely accepted view of music history as a succession of distinct eras (e.g.\ Romantic Period \cite{HistoryOfWesternMusic}, Rock Era \cite{Negus1996}, etc.).

Nonetheless, from the viewpoint of modelling the music creation process, however, it is difficult to represent the distributions of musical features of each time period simply by their means \cite{Mauch2015,Weiss2018}.
This is because there are typically multiple clusters (genres, modes, etc.) of musical styles coexisting concurrently in a society, and averaging the features of different clusters is not a logical approach in most practical cases.
For example, classical composers used both the major and minor scales, which can be represented as different probability distributions over pitches, but did not use the hybrid structure obtained by averaging these probability distributions.
To address this internal structure of musical feature distributions, previous studies \cite{Mauch2015,Weiss2018} have applied data-driven clustering methods and analyzed the evolution of the frequencies of the obtained clusters.
The clusters often corresponded to known genres or interpretable composition styles, and the revolutions in musical styles were often associated with the rise and decline of particular clusters (inter-cluster dynamics).
These results suggest that the distribution of music creation styles consists of concurrent and transient clusters, where the evolution of the distribution is caused by the dynamic changes of the clusters' relative frequencies.

The concurrent and transient cluster structure in music evolution raises further questions.
First, what internal (intra-cluster) dynamics do clusters have?
As jazz music, for example, is categorized into subgenres associated with certain time periods \cite{Gioia2021}, musical genres are often considered to have a hierarchical and dynamic structure.
In musicology, changes in the creation style of individual composers (e.g.\ Beethoven \cite{HistoryOfWesternMusic}) are also studied.
This question is important for evolutionary modelling because intra-cluster dynamics imply that a content-based artefact-level selective pressure is at work, in addition to the group-level selection implied by inter-cluster dynamics.
Second, to what extent can we predict the future distribution of music creation styles by incorporating the cluster dynamics?
To quantitatively study the relative contributions of the inter- and intra-cluster dynamics, it is necessary to develop a computational model that can infer and predict the inter- and intra-cluster dynamics from data.

To address these questions, we conduct evolutionary analyses of music data using statistical modelling methods.
We use two datasets of Japanese and US popular music, both containing songs appearing in top sales charts covering 50 years or more (the US and Japan had the largest music recording industries globally during the 2010s, according to the reports by the International Federation of the Phonographic Industry \cite{IFPI}).
We extract musical statistics from these data, apply clustering, and study the dynamic changes in the clusters' structure.
We observe some intra-cluster dynamics, such as changes in the clusters' means and variances.
We show that these dynamics can be described by a statistical model called the dynamic Dirichlet mixture model (DDMM), whereby the intra- and inter-cluster dynamics are translated into the time-evolving parameters of the DDMM in a unified manner.
We then formulate an evolutionary model for the DDMM parameters in which the fitness-based competitive dynamics of the clusters and those of musical elements within each cluster are incorporated.
Finally, we develop a method for inferring and predicting these parameters and quantitatively examine the contributions of the inter- and intra-cluster dynamics to predict the future distribution of music creation styles.

Theremainder of the paper is structured as follows.
In Sec.~\ref{sec:DynamicsOfStyleDistribution}, we analyze Japanese popular music data and the dynamics of cluster structure.
In Sec.~\ref{sec:PredictiveModelling}, we develop our predictive model.
Sec.~\ref{sec:USPop} presents the results for the analysis of US popular music data.
We summarize the findings and discuss their implications in Sec.~\ref{sec:Discussion}.

\section{Dynamics of creation style distributions}
\label{sec:DynamicsOfStyleDistribution}

In this section, we analyze the cluster structure in Japanese popular music data, finding that some clusters exhibit notable intra-cluster dynamics. We also point out that major aspects of cluster dynamics can be described by the DDMM.

\subsection{Data and analyzed musical statistics}
\label{sec:DataAnalysisMethod}

We used a dataset of Japanese popular songs (J-pop dataset) that comprised 1399 songs in the top sales charts provided by Oricon Inc.; the charts were mainly based on the amount of phonograph records and CD sales.
The vocal melodies of all songs that ranked within the top 20 in the yearly charts between 1950 and 2019 were transcribed and encoded in the MusicXML format (except for one song whose audio file was not accessible).
For computational analysis, each melody was represented as a sequence of pairs of integers, one representing the pitches and the other representing the onset times of musical notes.
The pitch of a note was represented in units of semitones, and all songs were transposed to the C major or A minor key prior to the analysis.
The onset time of a note was represented as an integer $b\in\{0,1,\ldots,47\}$ corresponding to its relative position in a bar (called the metrical position).
For example, $b=0$ indicates the downbeat position, and in 4/4 time, $b=12$ indicates the second beat position (we fixed the temporal resolution such that a bar has $48$ units).
\begin{figure*}
\centering
\includegraphics[width=1.9\columnwidth]{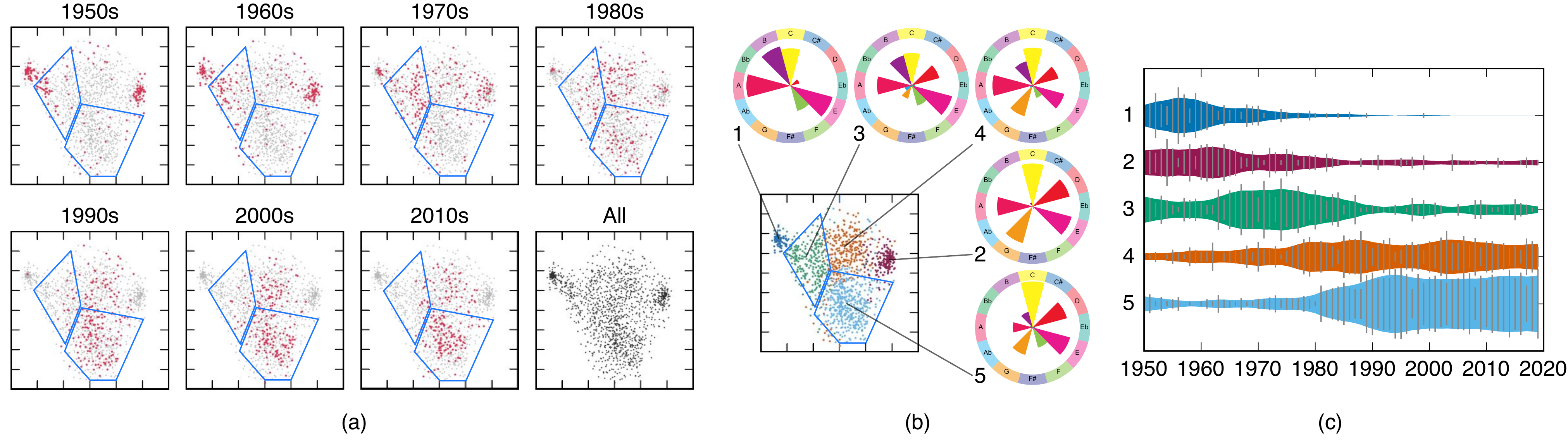}
\caption{The evolutions of the distribution and cluster structure of the pitch statistics in the J-pop data. (a) The two-dimensional visualization of the data distribution. (b) Result of the cluster analysis (the figures in circles represent the pitch-class distributions). (c) The evolution of the relative frequencies of the clusters.}
\label{fig_JPop_Epc_Cluster}
\end{figure*}

For musical statistics, we used the pitch and rhythm bigram probabilities to represent the creation style of each song.
These bigram probabilities are equivalent to the transition probabilities of Markov models and can describe different musical scales and rhythmic modes \cite{Ames1987}.
To obtain pitch bigrams, we used the extended pitch class representation, in which a transition from pitch $p'$ to pitch $p$ is represented as a bigram $(q',q)$ ($0\leq q'\leq11$ and $0\leq q\leq35$) given by $q'\equiv p'$ (mod 12) and $q\equiv q'+p-p'$ (mod 36).
Thereby, we obtained a representation independent of the pitch range and capable of discriminating pitch intervals between $-17$ and $17$ semitones.
Thus, the pitch bigram probabilities had $12\times36$ elements.
The rhythm bigram probabilities were calculated from the frequencies of the bigrams of metrical positions; they had $48\times 48$ elements.
Although we need both of the probabilities for pitches and metrical positions to construct a statistical model for melody generation, we treated them separately in the following analysis to gain a clearer interpretation of the results.

After representing each song $n$ with a tuple $(\bm\theta_n,t_n)$ of the probability vector $\bm\theta_n$ of its pitch/rhythm statistics and its created time $t_n$, we applied a clustering method based on the discrete distribution mixture model.
Specifically, we used the expectation-maximization (EM) algorithm \cite{Bishop2006} with random initialization to train $K$ sets of discrete distributions representing the clusters.
After convergence, each song was assigned to the cluster with the maximal likelihood.
Time information was not used in the clustering process.

\subsection{Qualitative analysis of dynamics of creation style distributions}

%
\begin{figure*}
\centering
\includegraphics[width=1.9\columnwidth]{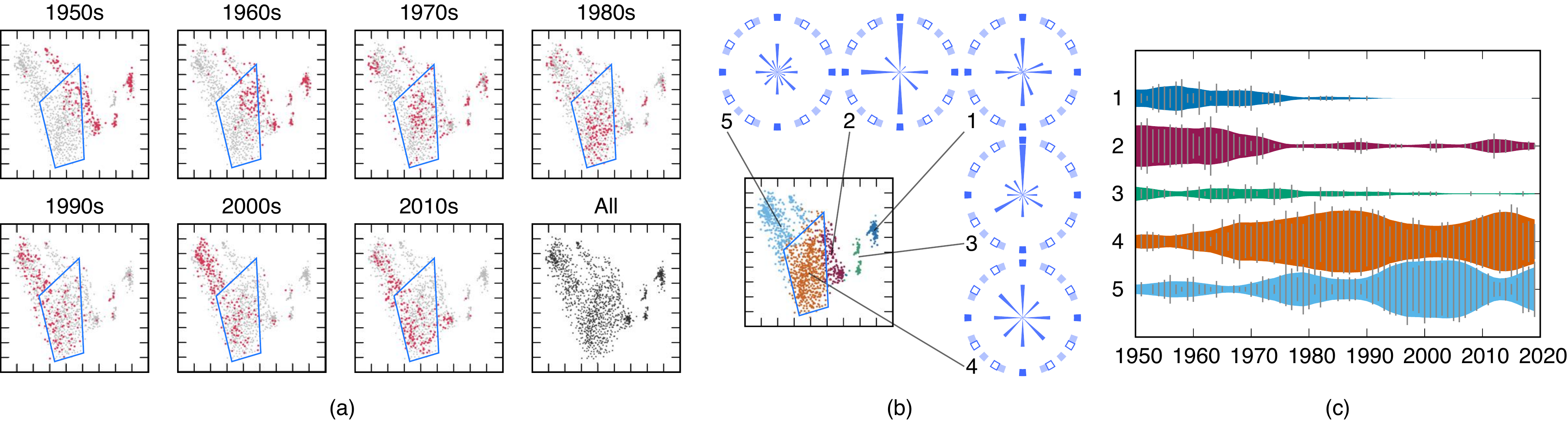}
\caption{The evolutions of the distribution and cluster structure of the rhythm statistics in the J-pop data. (a) The two-dimensional visualization of the data distribution. (b) Result of the cluster analysis (the figures in circles represent the distributions of metrical positions). (c) The evolution of the relative frequencies of the clusters.}
\label{fig_JPop_Met_Cluster}
\end{figure*}
To obtain an intuitive overview of the creation style distribution dynamics, we represented the probability vectors of songs in a two-dimensional space using the t-distributed stochastic neighbour embedding (t-SNE) visualization method \cite{Maaten2008}.
For this visualization, we first calculated the Jensen-Shannon (JS) divergence between each pair of probability vectors, which measures their ``distance''.
Then, the t-SNE method was used to map the probability vectors into the visualization space where the distances were retained as much as possible.

Before discussing the dynamics, we first discuss the cluster structure of the creation style distributions using the obtained visualization.
From the results for the pitch statistics in Fig.~\ref{fig_JPop_Epc_Cluster}(a), we can find a small cluster on the left and another small cluster on the right, as well as a large cluster spread in the central region.
Inside the large cluster, we can find some structures of density variations, but there are no clearly separated clusters.
The clustering method, assuming five classes ($K=5$), automatically identified the two small clusters and divided the large cluster into three smaller clusters (Fig.~\ref{fig_JPop_Epc_Cluster}(b)).
The average statistics for the clusters, represented as pitch-class probabilities, show that all five clusters are associated with distinguishable musical scales.
Cluster 1 is associated with the minor pentatonic scale, cluster 2 with the major pentatonic scale, cluster 3 with the minor hexatonic scale, cluster 4 with the minor diatonic scale, and cluster 5 with the major diatonic scale.
These musical scales, except the hexatonic scale, are well known to musicologists \cite{Stevens2008}.
From a macroscopic viewpoint, clusters 3, 4, and 5 form a family of diatonic scales.
These clusters in addition to the larger cluster that includes them form a hierarchical structure in the creation style distribution.

Similar observations can be drawn from the results for the rhythm statistics.
From Fig.~\ref{fig_JPop_Met_Cluster}(a), we find three clearly separated small clusters on the right and a large cluster with visible internal structures on the left.
Assuming five classes, the clustering method identified three clusters inside the large cluster and assigned two clusters to the separated clusters on the right (Fig.~\ref{fig_JPop_Epc_Cluster}(b)).
Again, we can musically interpret the five clusters using their average statistics, represented as metrical position probabilities.
Cluster 1 is associated with the dotted rhythm, cluster 2 with the 4 beat rhythm, cluster 3 with the triplet rhythm, cluster 4 with the 8 beat rhythm, and cluster 5 with the 16 beat rhythm.
This creation style distribution also has a hierarchical structure with a large cluster containing smaller clusters that overlap with each other.

These results provide some implications for the applicability of representing creation style distributions as a mixture of clusters.
On the one hand, representation using clusters can capture the internal structure of a creation style distribution more precisely than representing it as a simple (e.g.\ unimodal) distribution, and to some extent, it can reproduce style classifications studied by musicologists.
On the other hand, the clusters often overlap significantly and possibly form a hierarchical structure, which invalidates the notion of definitive clusters.
Despite these limitations, cluster representation is still effective for approximating the complex structure of creation style distributions using simple and tractable mathematical models and interpreting the results.

We now discuss the dynamics of creation style distributions.
From Figs.~\ref{fig_JPop_Epc_Cluster}(a) and \ref{fig_JPop_Met_Cluster}(a), we find drastic changes in the creation style distributions for both statistics.
For the pitch statistics, most songs were located in the left-most or right-most regions in the 1950s and 1960s, and in the central region in the 1990s and later.
This overall change is clearly visible in the evolution of cluster frequencies, which shows the dominance of clusters 1 and 2 in the early years and the dominance of clusters 4 and 5 in later years (Fig.~\ref{fig_JPop_Epc_Cluster}(c)).
Similarly, for rhythm statistics, most songs were located in the upper-right region in the panels for the 1950s and 1960s, and in the lower-left region in the panels for the 2000s and later.
The evolution of cluster frequencies in Fig.~\ref{fig_JPop_Met_Cluster}(c) also reflects this overall transition.
Therefore, the major transitions in the creation style distributions of Japanese popular songs can be captured by the inter-cluster dynamics, similar to the observations made in the evolution of US popular music \cite{Mauch2015} and Western classical music \cite{Weiss2018} (see also Sec.~\ref{sec:USPop}).

We also observed some dynamics within the clusters.
First, the distribution of the pitch statistics in cluster 5 reduced its variance from the 1990s to the 2010s.
This contraction of the cluster is ascribed to the movement towards market concentration during the period: 103 distinct artists appeared on the chart in the 1990s, whereas only 76 (resp.\ 32) distinct artists appeared in the 2000s (resp.\ the 2010s).
Second, the distribution of pitch statistics in cluster 3 exhibits a significant shift of a concentrated region from the 1950s to the 1980s.
This reflects the movement of musical preferences from the minor pentatonic scale to the minor diatonic scale in songs that appear on the chart.
A similar shift of a concentrated region can be observed in cluster 4 of the rhythm statistics between the 1960s and the 1980s.
This shift represents the movement towards widely utilizing syncopated rhythms (see SM).

These observations show that the creation style distributions exhibit dynamical modes that cannot be purely represented as changes in the frequencies of clusters having stationary distributions.
To capture the collective nature of a contraction or shift of a cluster, we should consider the intra-cluster dynamics, or equivalently, represent the cluster by its smaller components and consider their coherent frequency dynamics that also depend on their content features.
Therefore, the present analysis indicates that both inter- and intra-cluster dynamics are relevant for an effective modelling of the evolution of music creation style distributions.

\subsection{Quantitative analysis of intra-cluster dynamics using Dirichlet distribution}
\label{sec:IntraClusterDynamicsAndDirichletModel}

While the qualitative analysis in the previous section was intuitive, the two-dimensional visualization should be seen as an approximate picture of the distribution of creation styles, which were in fact represented as high-dimensional probability vectors.
To quantitatively analyze intra-cluster dynamics, we mathematically formulate clusters in the space of probability vectors.
Note that a standard statistical formulation uses the Gaussian distribution to represent a cluster in the Euclidean space \cite{Bishop2006}.
The Gaussian distribution has parameters that can represent the basic properties of a cluster (the mean vector represents the cluster's centre while the covariance matrix represents the cluster's size and direction) and is well-defined in the Euclidean space.
By extending this idea to the case of our concern, we can use the Dirichlet distribution, which is commonly used in Bayesian statistics to represent a cluster in the space of probability vectors \cite{Bishop2006}.
It is also known that the Dirichlet distribution can approximately reproduce the distribution of the Shannon entropies for rhythm statistics in popular music melodies \cite{Nakamura2021IS}.

For $D$-dimensional probability vectors $\bm\theta=(\theta_i)_{i=1}^D$ ($\sum_i\theta_i=1$), the Dirichlet distribution is defined as
\begin{equation}
{\rm Dir}(\bm\theta;\alpha,\bm\mu)=\Gamma(\alpha)\prod_{i=1}^D\frac{\theta_i^{\alpha\mu_i-1}}{\Gamma(\alpha\mu_i)},
\label{eq:DirichletModel}
\end{equation}
where the parameter $\alpha>0$ represents the concentration and the probability vector $\bm\mu=(\mu_i)_{i=1}^D$ is the mean (or base) distribution.
A Dirichlet distribution is a probability distribution of probability vectors, and the component-wise mean and variance are given as
$E(\theta_i)=\mu_i$ and $V(\theta_i)=\mu_i(1-\mu_i)/(\alpha+1)$, respectively.
Thus, the mean $\bm\mu$ can be interpreted as the centre of the cluster, and the concentration $\alpha$ is a parameter related to the cluster's size (the variance is inversely proportional to $\alpha+1$).
Given a set of probability vectors assigned to a cluster, we can estimate these parameters for the cluster using the maximum likelihood estimation method \cite{Minka2000}.
\begin{figure}
\centering
\includegraphics[width=1\columnwidth]{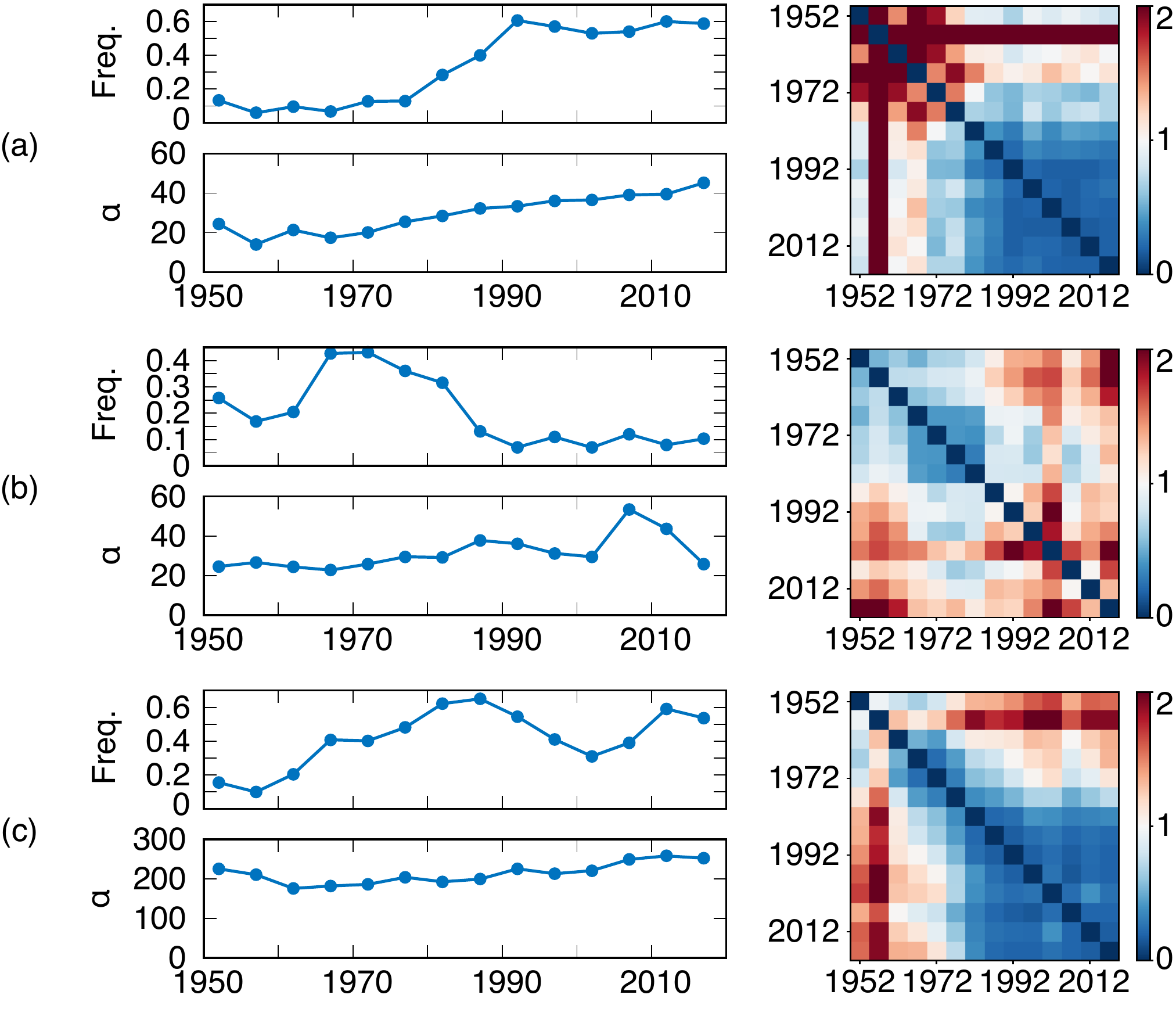}
\caption{The evolutions of the relative frequency, concentration, and mean distribution for (a) cluster 5 and (b) cluster 3 of the pitch statistics in Fig.~\ref{fig_JPop_Epc_Cluster} and for (c) cluster 4 of the rhythm statistics in Fig.~\ref{fig_JPop_Met_Cluster}. The evolution of the mean distribution is represented by the SKL divergences between time slice pairs on the right.}
\label{fig_IntraclusterAnalysis}
\end{figure}

Fig.~\ref{fig_IntraclusterAnalysis} shows the evolution of the estimated parameters of the Dirichlet distributions for some clusters computed after the data samples were divided into time slices.
To quantify the temporal differences in the mean distributions, we used the symmetric Kullback-Leibler (SKL) divergence, which is a distance measure for general probability distributions.
Fig.~\ref{fig_IntraclusterAnalysis}(a) illustrates the evolution of cluster 5 of the pitch statistics; the contraction of the cluster from the 1990s to the 2010s is represented as an increase in the concentration parameter.
Note that the SKL divergences of the mean distributions between any pair of time slices remained far below unity during this time period, indicating that the shift of the cluster centre was relatively small.
Fig.~\ref{fig_IntraclusterAnalysis}(b) illustrates the evolution of cluster 3 of the pitch statistics; the shift of the cluster centre from the 1950s to the 1980s is reflected in the relatively large off-diagonal elements in the SKL divergence matrix during this period.
Similarly, in Fig.~\ref{fig_IntraclusterAnalysis}(c), which illustrates the evolution of cluster 4 of the rhythm statistics, the off-diagonal elements in the SKL divergence matrix for the period between the 1960s and the 1980s take considerably larger values than those for the period between the 1990s and the 2010s, quantitatively demonstrating that a shift of the cluster centre occurred in the early years.

These results show that the Dirichlet distribution can be used to quantitatively analyze the intra-cluster dynamics of the creation styles.
In particular, a contraction (and possibly expansion) and shift of the centre of a cluster can be described as evolutions of the concentration and mean distribution, respectively.
Moreover, as shown in the next section, together with a dynamical model for the parameter evolution, the Dirichlet distribution can be used as a statistical model for predicting the future creation style distribution.
This formulation also allows us to evaluate the predictive accuracy by measuring the likelihood, following the general statistical framework.

\section{Predictive modelling}
\label{sec:PredictiveModelling}

\subsection{Dynamic Dirichlet mixture model (DDMM)}

In Sec.~\ref{sec:DynamicsOfStyleDistribution}, we found that both inter- and intra-cluster dynamics are important aspects of the evolution of music creation style distributions.
Here, we construct a model incorporating these dynamics to predict future creation style distributions and examine the relevance of intra-cluster dynamics for predictive modelling.
As explained in Sec.~\ref{sec:IntraClusterDynamicsAndDirichletModel}, the clusters of creation styles in the space of probability vectors can be described by Dirichlet distributions.
Thus, the distribution of creation styles composed of multiple clusters can be described by a Dirichlet mixture model (DMM) that is defined as
\begin{equation}
P(\bm\theta,t)=\sum_{k=1}^K\pi_k(t)\,{\rm Dir}(\bm\theta;\alpha_k(t),\bm\mu_k(t)),
\label{eq:DynDMM}
\end{equation}
where $\bm\theta$ represents a probability vector as in Eq.~(\ref{eq:DirichletModel}), $k$ indexes clusters, and $K$ denotes the number of clusters.
The quantity $\pi_k$ (called the mixture probability) represents the relative frequency of cluster $k$ (we have $\sum_k\pi_k=1$).

In Eq.~(\ref{eq:DynDMM}), the dynamics of the creation style distribution are decomposed into the time evolutions of the parameters on the right-hand side, which is the basic assumption of the present model.
The inter-cluster dynamics are incorporated in the time-dependent mixture probabilities $\pi_k(t)$, representing the changes in the cluster frequencies.
The intra-cluster dynamics are incorporated in the time-dependent means $\bm\mu_k(t)$, representing the shifts of the cluster centres, and the time-dependent concentration parameters $\alpha_k(t)$, representing the contractions and expansions of the clusters.
Hereafter, the model in Eq.~(\ref{eq:DynDMM}) is called the dynamic Dirichlet mixture model (DDMM).

To use the DDMM for predicting future creation style distributions, given data up to the present, two problems should be addressed.
The first is the parameter estimation problem, that is, estimating the parameters $\pi_k(t)$, $\alpha_k(t)$, and $\bm\mu_k(t)$ at each time point up to the present that optimally fit the given data.
The second is the parameter prediction problem, that is, predicting future values of the parameters given the parameter values up to the present.
We discuss the first problem in Sec.~\ref{sec:ParameterEstimation} and the second problem in Sec.~\ref{sec:ParameterPrediction}.
As a problem setup, we consider the situation in which the data samples are assigned to one of the clusters, as in Sec.~\ref{sec:DynamicsOfStyleDistribution}.

\subsection{Parameter estimation problem}
\label{sec:ParameterEstimation}

A simple solution to the parameter estimation problem is to learn the parameters for each year (or other units of time slice), $t$, using the data samples created in that year.
The mixture probabilities $\pi_k(t)$ and mean distributions $\bm\mu_k(t)$ can be obtained by taking the sample averages, and the concentration $\alpha_k(t)$ can also be determined using the maximum likelihood estimation method \cite{Minka2000}.
However, this simple method suffers from the data sparseness problem when the available data are not sufficient and the number of analyzed statistics is large, as is the case for the present analysis.

To address this problem, one approach is to use all data samples created by time $t$ with certain weighting factors to learn the parameters.
Using the standard exponential decay factor, this weighted average method can be formulated as follows.
The mixture probabilities and mean distributions can be obtained as
\begin{align}
\pi^{\rm WA}_k(t)&=C(t)^{-1}\sum_{s=-\infty}^t e^{(s-t)/\tau}\,\pi^{\rm S}_k(s),
\\
\bm\mu^{\rm WA}_k(t)&=C(t)^{-1}\sum^t_{s=-\infty} e^{(s-t)/\tau}\,\bm\mu^{\rm S}_k(s),
\end{align}
where $\pi^{\rm S}_k(s)$ and $\bm\mu^{\rm S}_k(s)$ are the values obtained using the aforementioned simple method, $e^{(s-t)/\tau}$ is the decay factor, $\tau$ is the time constant, and $C(t)=\sum^t_{s=-\infty} e^{(s-t)/\tau}$ is a normalization factor.
For the concentration parameter, we can use the log-likelihood obtained from a single-time log-likelihood $L_k^{\rm S}(\alpha;s)$ calculated from Eq.~(\ref{eq:DirichletModel}) for cluster $k$, as
\begin{align}
L_k^{\rm WA}(\alpha;t)=C(t)^{-1}\sum^t_{s=-\infty} e^{(s-t)/\tau}\,L_k^{\rm S}(\alpha;s),
\end{align}
and then estimate its value as $\alpha^{\rm WA}_k(t)={\rm argmax}_\alpha\,L_k^{\rm WA}(\alpha;t)$.

\begin{figure}
\centering
\includegraphics[width=0.95\columnwidth]{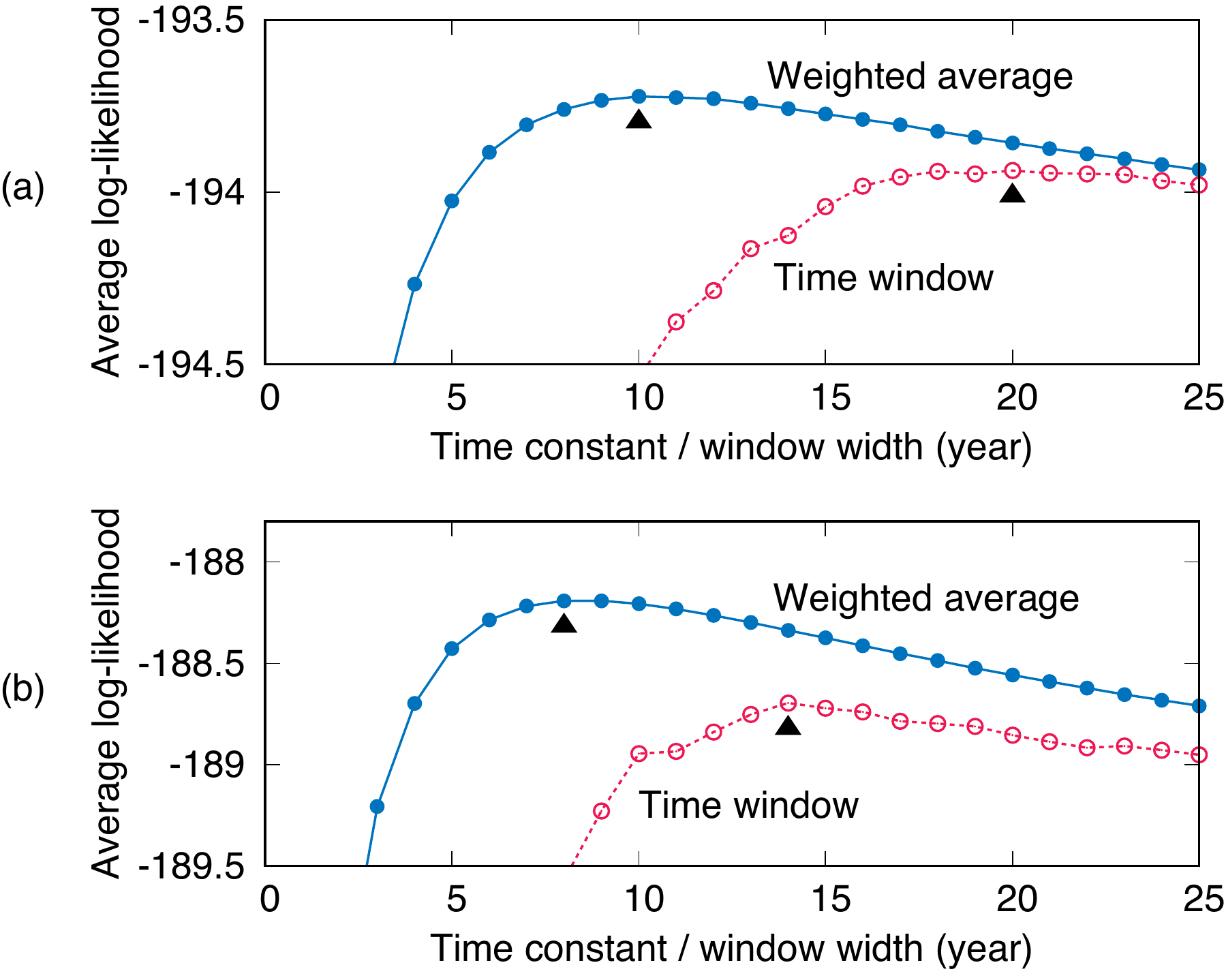}
\caption{Log-likelihood of the DMMs ($K=5$) estimated by the weighted average method and time window method, for the (a) pitch statistics and (b) rhythm statistics of the J-pop dataset.}
\label{fig_InstantPrediction_OptTimeConstant}
\end{figure}

A suitable value for the time constant $\tau$ can be determined from the data by optimizing the likelihood of the estimated DMM.
Fig.~\ref{fig_InstantPrediction_OptTimeConstant} shows the how the likelihood depends on the value of $\tau$.
To obtain this result, we set the referential year (tentative ``present year'' for the analysis) $t_0$ to each value in the range $[1960,2018]$.
Then, we estimated the DMM parameters, and calculated the likelihood of the data created in year $t_0+1$ (assuming that the creation style distribution would not change significantly in one year).
The likelihood was optimal at $\tau=10$ for the pitch statistics and $\tau=8$ for the rhythm statistics.
We used these values in the following analysis.
For comparison, the result of estimating the DMM parameters using data samples in a time window of various widths (using a one-year width is the same as the simple method) is also shown in Fig.~\ref{fig_InstantPrediction_OptTimeConstant}, demonstrating the superiority of the weighted average method.

\subsection{Parameter prediction problem}
\label{sec:ParameterPrediction}

To solve the parameter prediction problem, we construct models for the inter- and intra-cluster dynamics.
These dynamics are considered to be a result of the creators' preferences to produce songs in a particular style and the public's preferences to appreciate songs in certain styles.
Without these preferences, creation styles would be reproduced simply by learning from random samples of past songs.
Then, drawing from the ideas of evolutionary biology \cite{Crow1970,Nowak2006}, the inter-cluster dynamics can be modelled as a reproduction process with competition, where the reproducing organisms correspond to clusters and the fitness (reproduction rate) is determined by the preferences of the creators and the public.
The intra-cluster dynamics can be modelled by a similar reproduction process that includes competitions among constituent elements of musical pieces, where the fitness represents the preferences of the creators and the public with respect to musical contents.

To use these evolutionary models for the prediction problem, it is necessary to estimate fitness from past data.
Since such evolutionary processes are usually stochastic, while we can only observe a finite (and often small) number of data samples, statistical noise should be considered for the fitness estimation.
In addition, as musical preferences themselves are also likely to change over time \cite{Interiano2018}, a model that allows extensions to incorporate the evolution of fitness is needed.
In the following section, we formulate a fitness-based model within a statistical framework that allows a robust estimation of fitness and extensions to dynamically changing fitness.

We first present a model for predicting the evolution of mixture probabilities $\pi_k(t)$.
The basic equation of fitness-based evolutionary dynamics \cite{Crow1970} can be represented as
\begin{equation}
\pi_k(t+1)=w_k\pi_k(t)/\bar{w},
\end{equation}
or
\begin{equation}
{\rm ln}\,\pi_k(t+1)={\rm ln}\,\pi_k(t)+{\rm ln}(w_k/\bar{w}),
\end{equation}
where $w_k$ denotes the fitness (reproduction rate) for cluster $k$, and the average fitness $\bar{w}$ is defined as $\bar{w}=\sum_\ell w_\ell\pi_\ell(t)$.
The second equation, which is a simple arrangement of the first equation, can be viewed as the equation of motion for the log probabilities, where the logarithm of the relative fitness $w_k/\bar{w}$ corresponds to the velocity.

There are two points to be discussed before moving forward.
First, although the relative fitness depends on all $\pi_\ell(t)$ through $\bar{w}$, the effect of the other $\pi_\ell(t)$ on the dynamics of $\pi_k(t)$ is small if the speed of evolution is slow.
In this approximation of slow evolution, which we adopt in this study, the dynamics of $\pi_k(t)$ are decoupled from each other and can be treated independently (until they are finally normalized after solving the dynamics).
Second, the above equations describe a model for a large population and do not include the statistical noise observed in a finite set of data samples.
If we write $x(t)={\rm ln}\,\pi_k(t)$ and $v(t)={\rm ln}(w_k/\bar{w})$, and represent the noise added to the observed $x(t+1)$ by $\epsilon(t)$, then the modified model is given as
\begin{equation}
x(t+1)=x(t)+v(t)+\epsilon(t).
\label{eq:EOM1}
\end{equation}
Specifically, $\epsilon(t)$ is described by a Gaussian noise that satisfies $\langle \epsilon(t)\rangle=0$ and $\langle\epsilon(t)\epsilon(s)\rangle=\sigma^2\delta_{ts}$.  
In the approximation of a constant fitness, $v(t)$ is independent of time.

We can extend the model to include the dynamics of the fitness by treating $v(t)$ as a time-dependent variable and adding a dynamical equation.
We obtain a natural model for $v(t)$ by introducing a variable $\eta(t)$ representing the temporal variation of $v(t)$ as
\begin{equation}
v(t+1)=v(t)+\eta(t),
\label{eq:EOM2}
\end{equation}
where $\eta(t)$ is also described by Gaussian noise.
In general, we can also introduce variables representing higher-order time derivatives and construct a general model (see Supplemental Material (SM)).
Eqs.~(\ref{eq:EOM1}) and (\ref{eq:EOM2}) (with possible additional equations) are known as the state-space model in statistics \cite{Bishop2006}.
An advantage of the present formulation is that we can apply the well-developed statistical theory of state-space models to estimate the fitness (or velocity $v(t)$) from the data and predict the future values of $x(t)$.
The details of this method are presented in SM.

The same model can be applied to predict the evolutions of the mean distributions $\mu_{ki}(t)$ and concentrations $\alpha_k(t)$.
In the case of mean distributions, fitness represents the preferences of the creators and the public with respect to constituent musical elements indexed by $i$.
As concentrations are not frequency parameters, there are no intuitive interpretations of the corresponding fitness.
Nevertheless, if the evolution of the concentrations results from dynamics where their velocities vary smoothly over time, a state-space model can serve as an approximating model for the dynamics.
Therefore, we apply such a model in the following analysis.

In a preliminary analysis, we found that using predictions that are more conservative than those directly obtained by the model often leads to higher accuracy.
More precisely, after the velocity is predicted by the model, it is reduced by a factor called the fitness reduction factor before it is used for making predictions.
In our analysis, the reduction factor was optimized for each set of parameters (mixture probabilities, concentrations, and mean distributions) in the range $[0.01,1]$.
See SM for details.

\subsection{Evaluation of predictive accuracies}
\label{sec:EvaluationOfPredAccuracy}

To evaluate the predictive model for creation style distributions, we conducted two computational experiments.
In the first experiment, we evaluated the predictive accuracies of the DDMM parameters by component and compared them with other possible methods to examine the effectiveness of the fitness-based evolutionary model.
In the second experiment, we evaluated the overall predictive accuracy in terms of the likelihood of predicted distributions to examine the relevance of the predictive models for the component parameters and the dependence on prediction times.

In the first experiment (i.e.\ DDMM parameter prediction), we estimated the DDMM parameters using the music data created before and in a referential year $t_0$ by the method presented in Sec.~\ref{sec:ParameterEstimation}.
We then predicted the model parameters for later years using the method presented in Sec.~\ref{sec:ParameterPrediction}.
This prediction method will be referred to as the state-space evolutionary model (SSEM).
For comparison, we refer to a method that simply uses as predictions the model parameters estimated for the referential year as the `static model'.
We also compared the predictions made by the Gaussian process (GP) model \cite{GP}, which is a general-purpose regression method, and in particular, a generalization of the linear regression model.
We used a multi-kernel GP composed of a linear kernel, Gaussian kernel, Mat\'ern kernel with degree $\nu=5/2$, and a bias term.
When applying the SSEM and GP to predict the values of the mean distributions, we used these methods only for components with an average probability in the past data above a threshold of $0.01$ and used the static model for the other rare components.
This is because the probabilities of rare components are susceptible to large statistical noise and cannot be reliably estimated.
The weights and kernel parameters were optimized to predict each parameter (the GPy library \cite{gpy2014} was used in the analysis).
To evaluate the predictive ability, we computed the errors for each set of DDMM parameters.
We used the SKL divergence to measure the prediction errors for the mixture probabilities and mean distributions, and used the log-squared error for the concentration parameters.

\begin{figure}
\centering
\includegraphics[width=0.99\columnwidth]{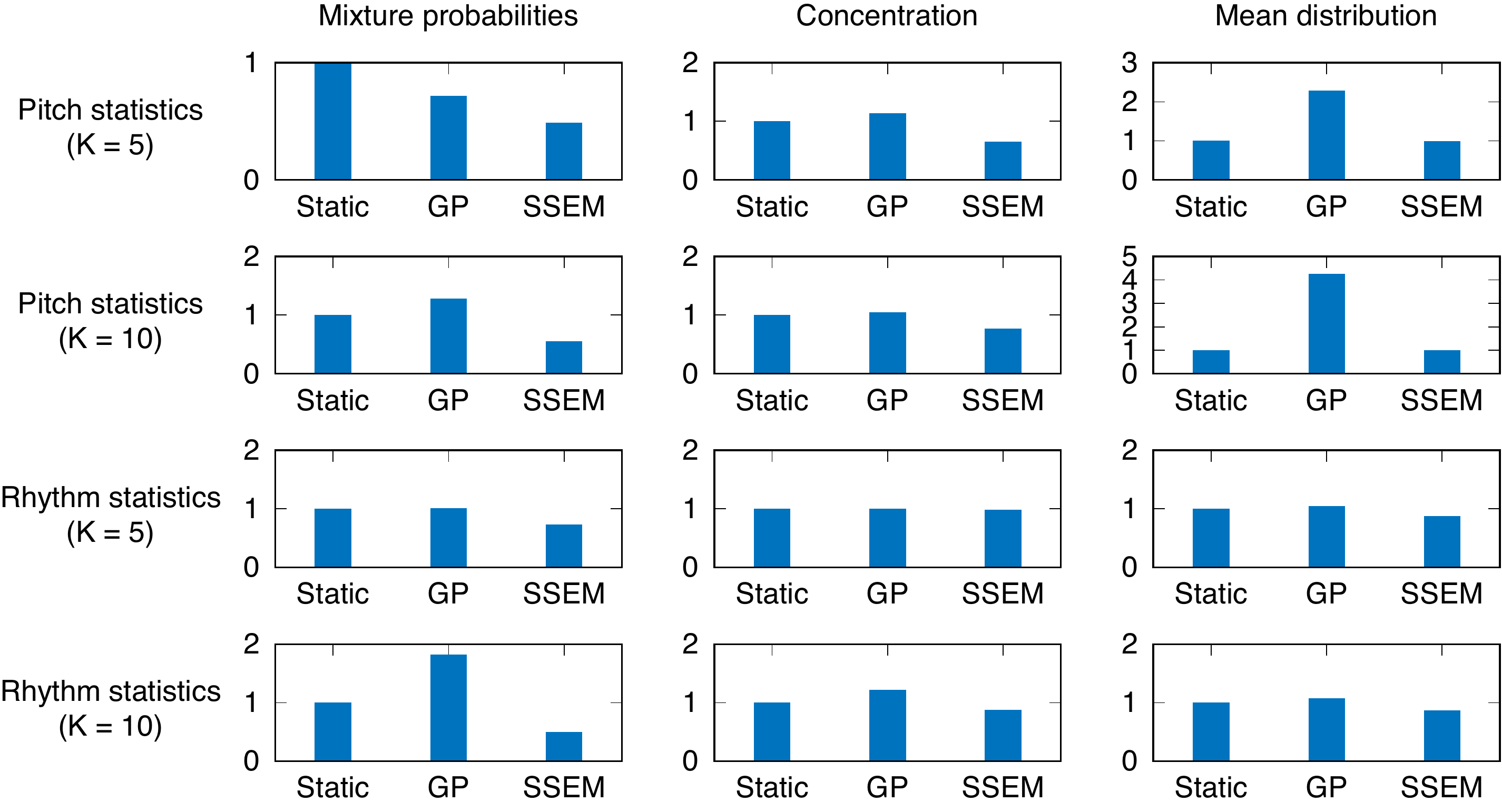}
\caption{Errors of DDMM parameter prediction by the static model, Gaussian process (GP), and the state-space evolutionary model (SSEM). All errors are normalized by the values for the static model.}
\label{fig_Summary_DDMMPrediction_JPop}
\end{figure}
The results are summarized in Fig.~\ref{fig_Summary_DDMMPrediction_JPop}, where the aforementioned three methods are compared for both pitch and rhythm statistics for the range $t_0\in[1969,2009]$.
For the mixture probabilities, the SSEM had significantly smaller errors than the static model in all cases.
The GP model had larger errors than the static model, except for the pitch statistics and the number of clusters $K=5$.
For the concentrations, the SSEM had smaller errors than the static model, except for the case of pitch statistics and $K=10$, where the errors for these models were similar.
The GP model had a similar or larger error than the static model for all cases.
For the mean distribution parameters, the SSEM and the static model had similar errors in the case of pitch statistics, while the former model had slightly smaller errors in the case of rhythm statistics.
The GP model had the largest errors in all cases.

Overall, the GP model often had larger prediction errors than the static model, indicating that the prediction of the DDMM parameters is a nontrivial problem.
In most cases, the SSEM can predict the parameters more accurately than the static model.
In particular, the reduction in error in the mixture probability predictions was considerably large, suggesting the effectiveness of describing the dynamics by a fitness-based evolutionary process with smoothly varying fitness values for the clusters.
However, the error reductions in the mean distribution parameters were relatively small, implying that the parameters follow more complicated dynamics.
\begin{figure*}
\centering
\includegraphics[width=1.9\columnwidth]{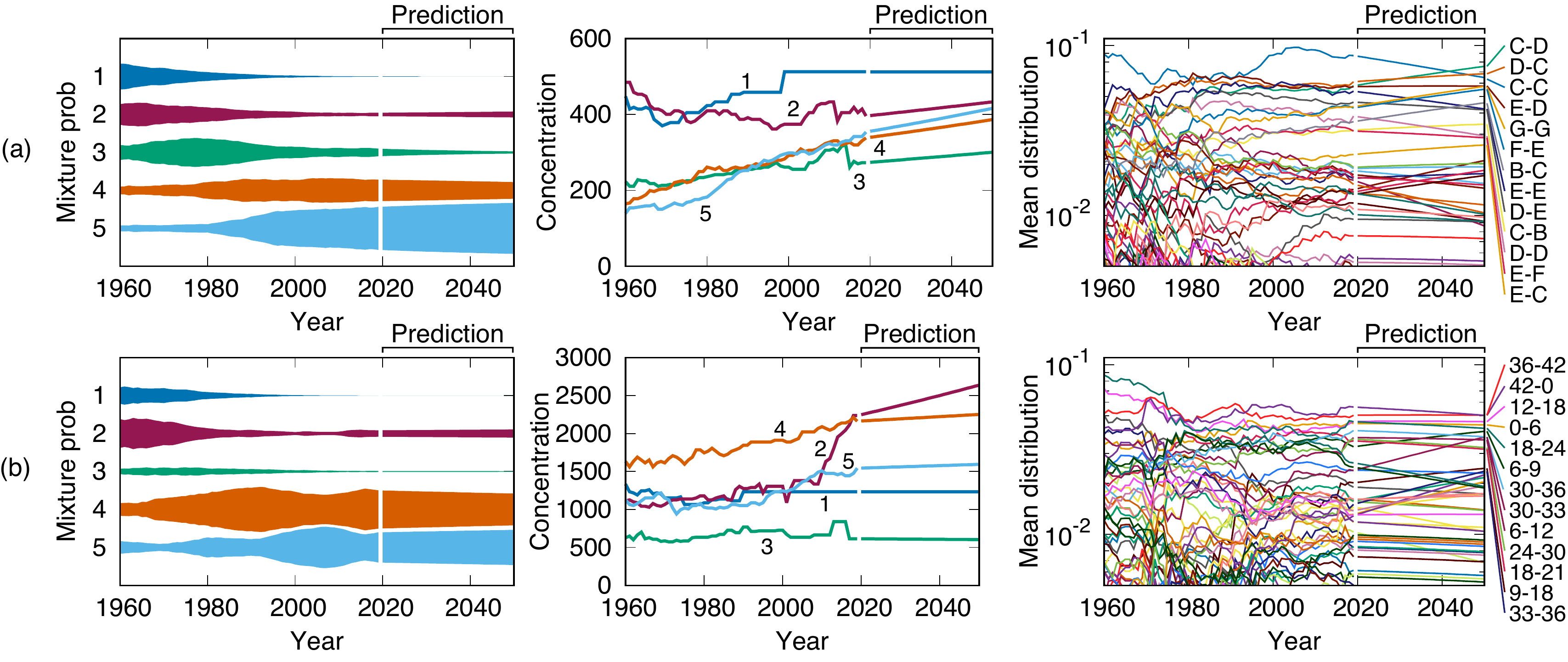}
\caption{The predicted evolution of creation style distributions of the (a) pitch and (b) rhythm statistics. The rightmost panel shows the mean distributions of the fifth cluster, and the labels on the right-hand side indicate the corresponding bigrams.}
\label{fig_Prediction_JPop}
\end{figure*}

The future values of the DDMM parameters predicted by the SSEM are shown in Fig.~\ref{fig_Prediction_JPop} for the case with $K=5$ clusters.
To make it easier to examine trends, the fitness reduction factor was set to unity for the mean distribution parameters to obtain these results.
For both pitch and rhythm statistics, it is predicted that the fifth cluster will continue to rise and become dominant.
For rhythm statistics, the second cluster is also predicted to rise in the near future.
The concentrations for all rising clusters are predicted to increase, indicating a decrease in the variabilities within individual clusters.

Focusing on the fifth cluster of the pitch statistics, some changes in the probabilities of the most frequent bigrams are predicted.
In particular, the probabilities of stepwise motions, such as (C,D), (F,E), and (B,C), are predicted to increase, whereas those of same-tone motions, such as (C,C), (E,E), and (D,D), are predicted to decrease.
For the fifth cluster of the rhythm statistics, the probabilities of currently the most frequent bigrams are predicted to eventually decrease, indicating a general uniformization of bigram probabilities.
Notable exceptions are bigrams with a 16th-note length, such as (6,9), (30,33), and (33,36), for which relatively rapid increases in the probabilities are predicted.
The observed trends in the to-be-dominant clusters of the pitch and rhythm statistics are consistent in reducing monotonic motions (same-tone or 8-beat rhythm).
To illustrate the effect of these parameter changes, some melodies generated by the mean distributions observed in 2019 and those predicted in 2040 are given in SM.
Perceptually, however, it is difficult to clearly distinguish the styles of these melodies.

In the second experiment (i.e.\ prediction of the creation style distributions), we measured the likelihood of the probability vectors in a range of 20 years after a referential year $t_0$, using the parameters predicted by the SSEM.
For predicting the creation style distribution at year $t_p$ after $t_0$, we used the parameters predicted for year $t_p+\Delta_p$, where $\Delta_p$ is a `look-ahead' parameter that compensates for the effect of the weighted average smoothing.
The value of $\Delta_p$ was optimized to maximize the likelihood, resulting in values $\Delta_p=8$ and $15$ for the pitch and rhythm statistics, respectively.
For the same purpose, we also used increased values for concentrations because the creation style distribution of an individual year was expected to have a larger concentration (smaller variance) than the smoothed one.
For the analysis, an increasing factor of $5$ was used for all data.
For comparison, we measured the likelihood of the same data using the static model explained above.
We evaluated cases $K=5$ and $10$ for the number of clusters, and used the static model with a single cluster ($K=1$) as the baseline.

\begin{figure}
\centering
\includegraphics[width=1.\columnwidth]{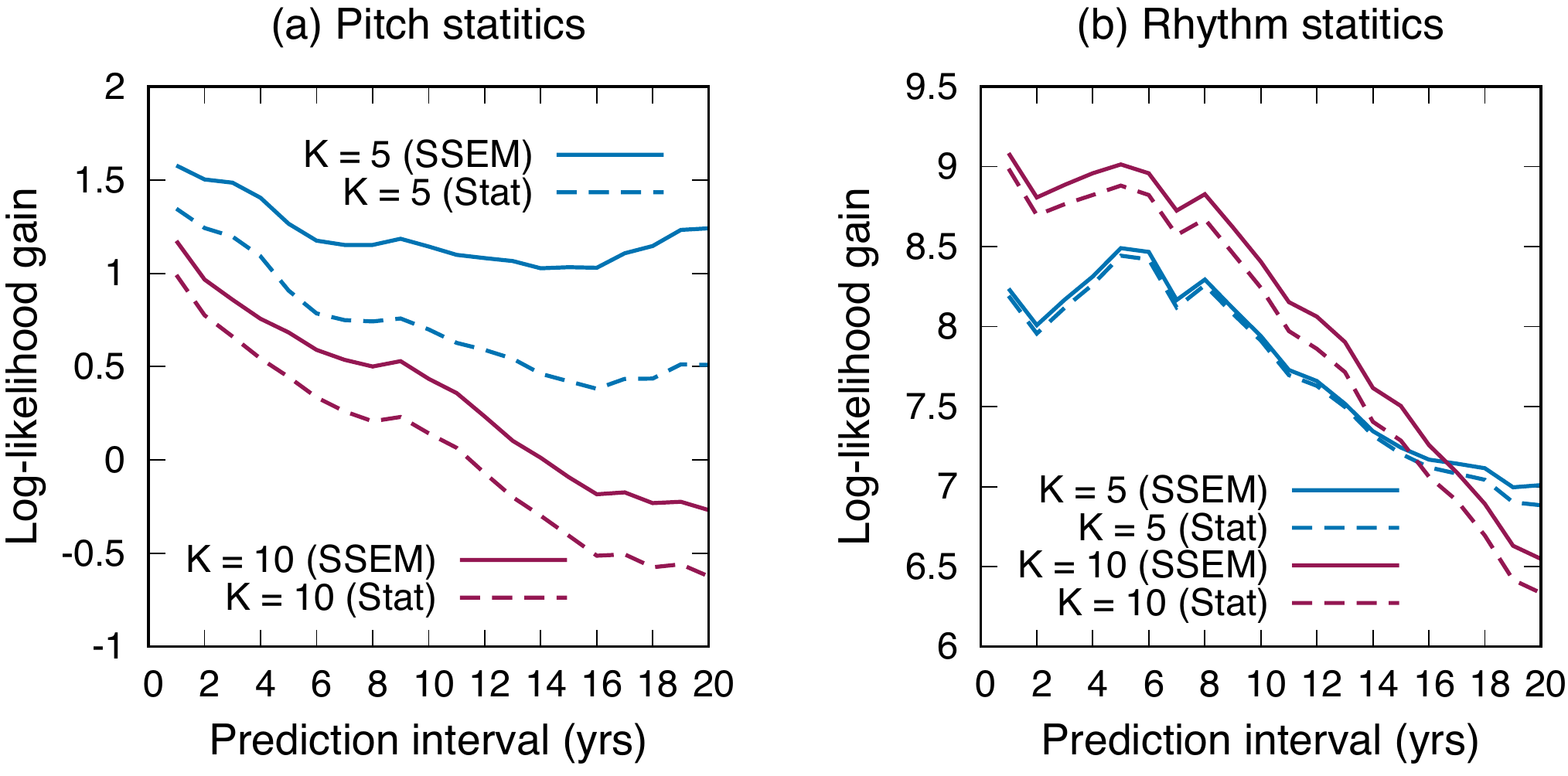}
\caption{The log-likelihood gains obtained by the SSEM and static model for the (a) pitch and (b) rhythm statistics.}
\label{fig_LogLikelihoodGain_JPop}
\end{figure}

Fig.~\ref{fig_LogLikelihoodGain_JPop} shows the dependence of the log-likelihood gains (compared to the $K=1$ static model) on the predictive intervals (which is defined as the year of prediction $-$ the referential year), averaged over all referential years in the range $t_0\in[1969,2009]$.
First, in all cases, the SSEM had a larger likelihood than the static model, indicating that the DDMM parameter prediction made by the SSEM can assist in predicting the distribution of the creation styles up to a prediction interval of at least 20 years.
Second, for both the pitch and rhythm statistics, the likelihood gains of the models with $K=10$ tended to decrease for longer prediction intervals and fell beneath the gains of the model with $K=1$ (pitch statistics) or $K=5$ (rhythm statistics) for predictive intervals of nearly 20 years.
This result indicates the difficulty of predicting the distribution increases for large numbers of clusters.
In addition to the possible effects of overfitting and data sparseness, the increase in the number of parameters that are difficult to predict might have led to the deterioration of the likelihood gains.

\begin{figure}
\centering
\includegraphics[width=1.\columnwidth]{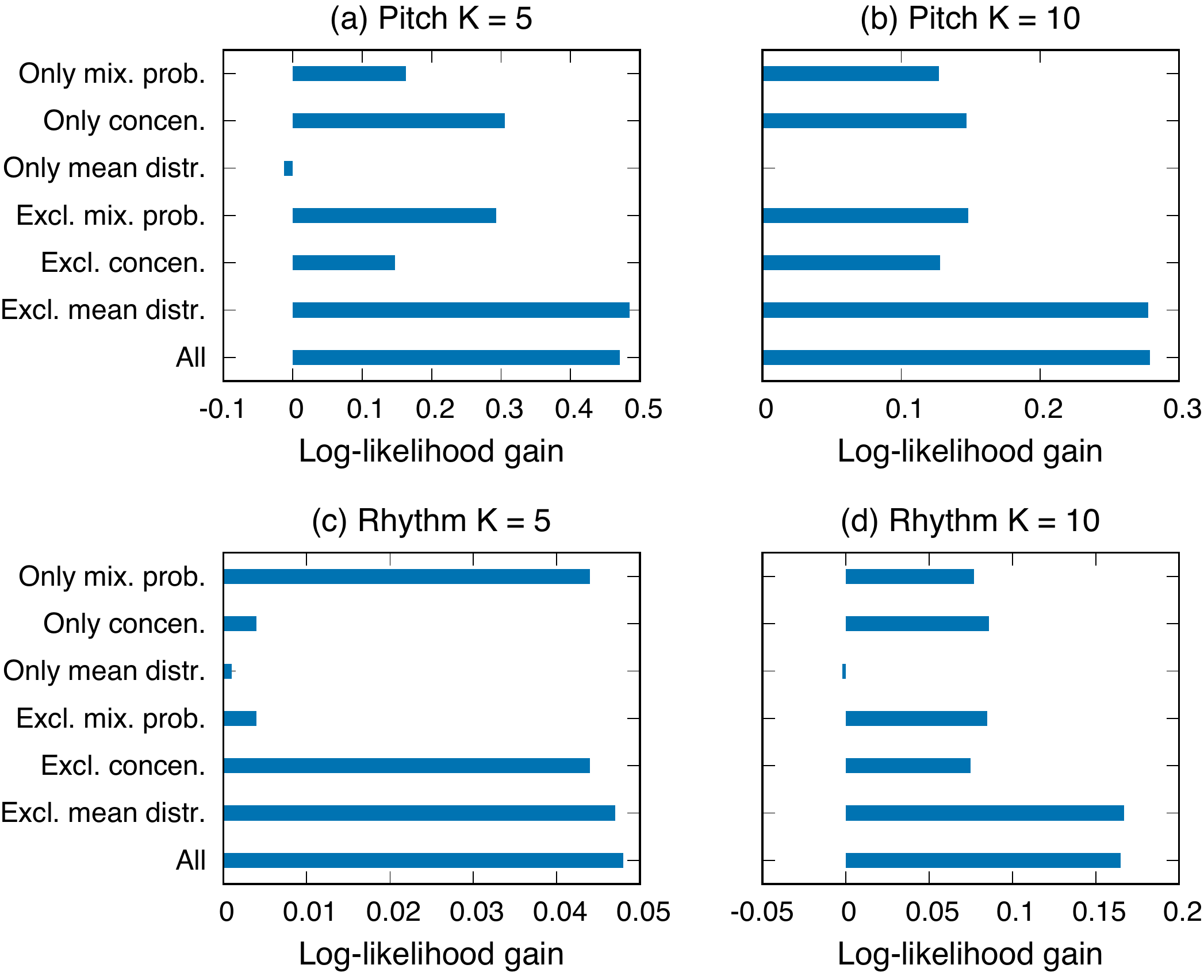}
\caption{The contributions of individual elements of the DDMM parameters to the log-likelihood gains. (a) Pitch statistics ($K=5$), (b) pitch statistics ($K=10$), (c) rhythm statistics ($K=5$), and (d) rhythm statistics ($K=10$).}
\label{fig_ElementwiseContributionsToLogLikelihoodGain_JPop}
\end{figure}

Finally, to examine the contributions of individual elements of the DDMM parameters to the likelihood gains, we compared the gains obtained by using the predictions by the SSEM for partial sets of the parameters and using the static model's predictions elsewhere.
The results in Fig.~\ref{fig_ElementwiseContributionsToLogLikelihoodGain_JPop} show that the SSEM's predictions of the mixture probabilities and concentrations contribute to the likelihood gains to some extent, and their contributions are often comparative.
This indicates that the predictions of both inter- and intra-cluster dynamics have positive effects on likelihood gains.
However, the SSEM's predictions of the mean distribution parameters have small and sometimes even negative effects.
Again, this is mainly due to the difficulty of predicting these parameters.
%

\section{Case of US popular music}
\label{sec:USPop}

%
\begin{figure*}
\centering
\includegraphics[width=1.7\columnwidth]{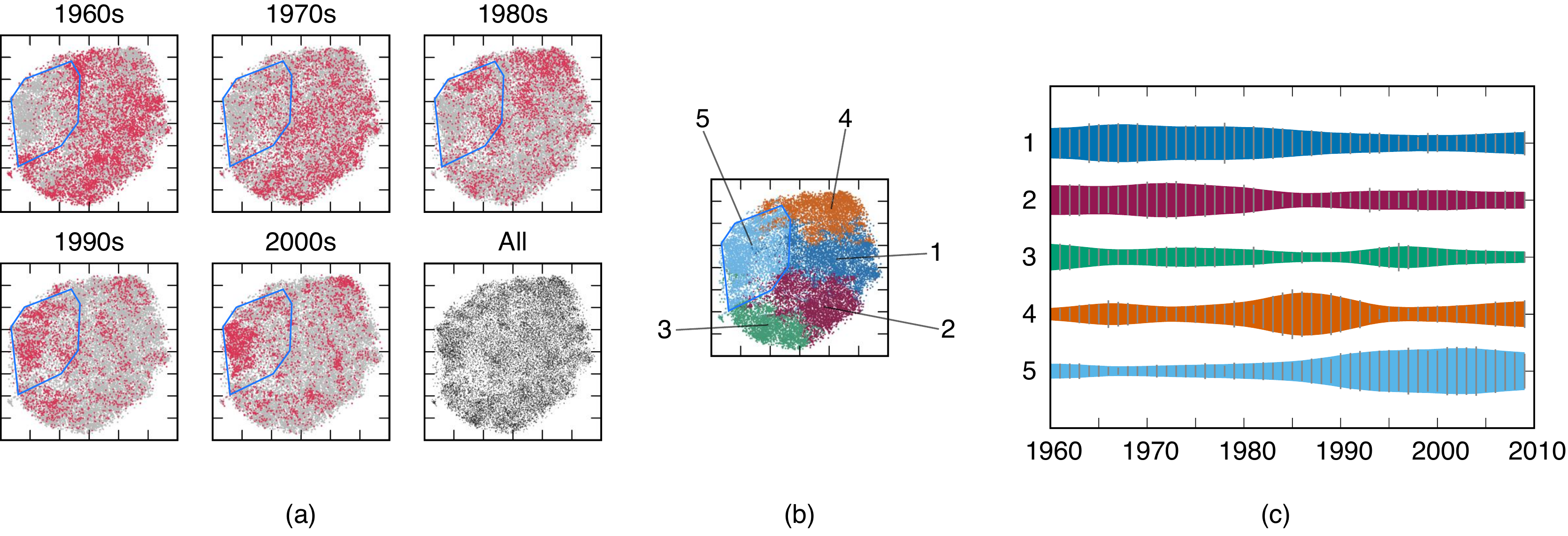}
\caption{The evolutions of the distribution and cluster structure of the timbre statistics in the US-pop data. (a) The two-dimensional visualization of the data distribution. (b) Result of the cluster analysis. (c) The evolution of the relative frequencies of the clusters.}
\label{fig_USPop_TLex_Cluster}
\end{figure*}
\begin{figure*}
\centering
\includegraphics[width=1.7\columnwidth]{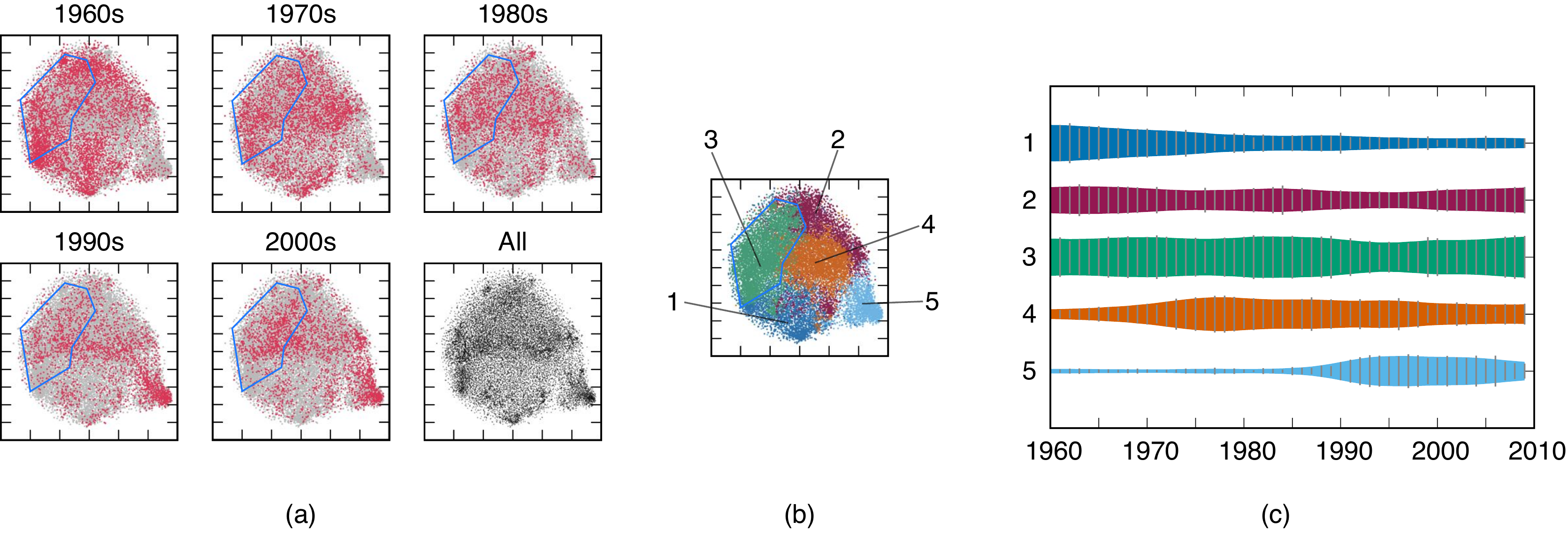}
\caption{The evolutions of the distribution and cluster structure of the harmony statistics in the US-pop data. (a) The two-dimensional visualization of the data distribution. (b) Result of the cluster analysis. (c) The evolution of the relative frequencies of the clusters.}
\label{fig_USPop_HLex_Cluster}
\end{figure*}
In this section, we analyze the US popular music data (US-pop dataset), developed by Mauch et al.~\cite{Mauch2015}, using the methods described in Secs.~\ref{sec:DynamicsOfStyleDistribution} and \ref{sec:PredictiveModelling}.
The dataset comprises 17\,094 songs that appeared in the Billboard Hot 100 charts between 1960 and 2009.
The audio clips of the songs were analyzed to extracte timbral and harmonic features.
We analyzed the lexicalized representations of these features obtained in \cite{Mauch2015}, where each song was represented by sequences composed of 35 `T-lexicons' and 193 `H-lexicons', respectively, and used their unigram probabilities as timbre and harmony statistics.

The two-dimensional visualization and clustering with $K=5$ classes in Figs.~\ref{fig_USPop_TLex_Cluster} and \ref{fig_USPop_HLex_Cluster} show that the creation style distributions changed considerably during the analysis period, and that the cluster structure was not very clear in these data.
While the overall changes can be represented as the evolution of relative frequencies of the clusters, we can also observe intra-cluster dynamics in both data.
For example, cluster 5 of the timbre statistics contracted between the 1990s and the 2000s, while cluster 3 of the harmony statistics shifted its density distribution between the 1960s and the 1990s.

\begin{figure}
\centering
\includegraphics[width=0.99\columnwidth]{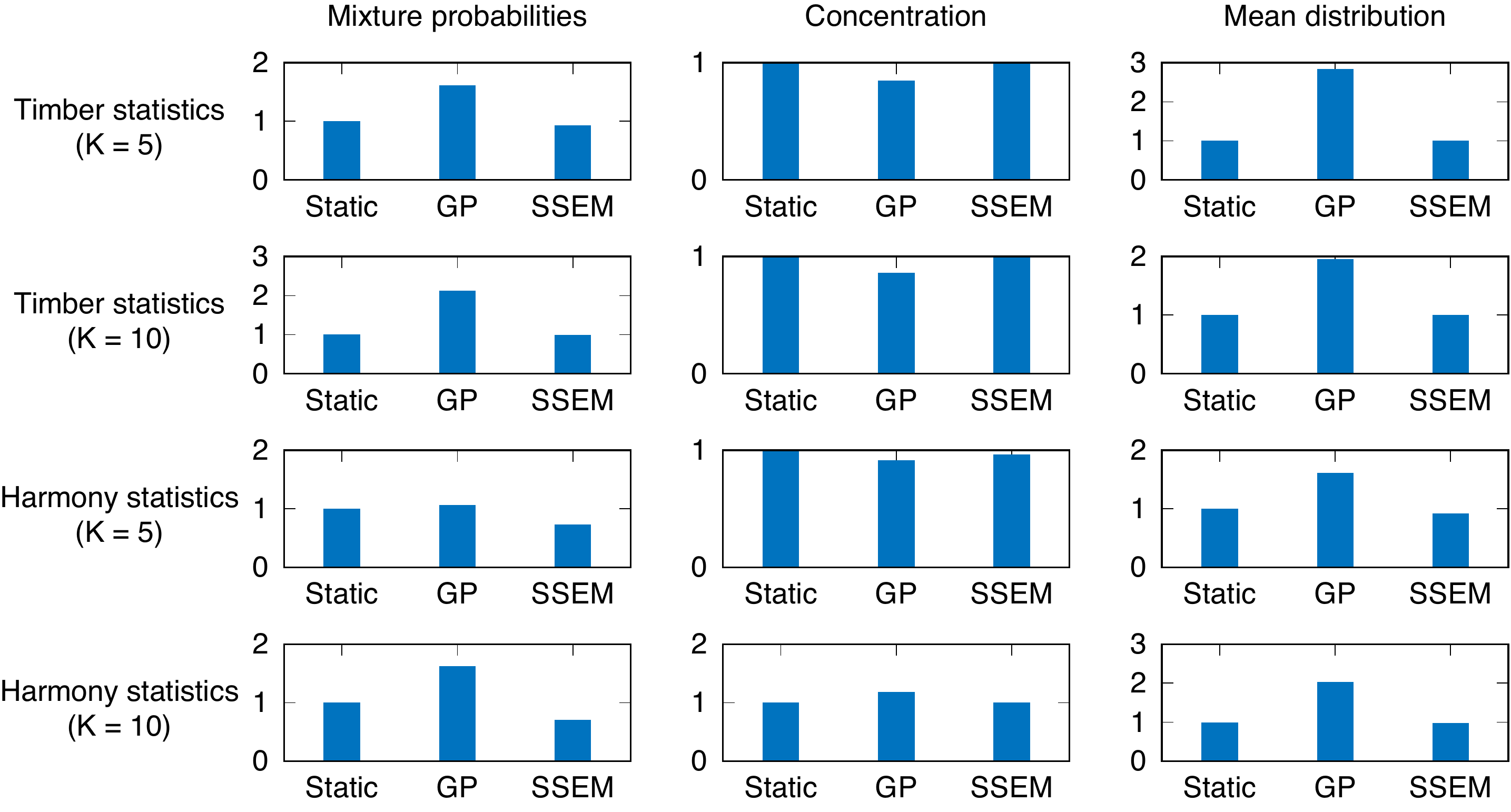}
\caption{Errors of the DDMM parameter prediction by the static model, Gaussian process (GP), and the state-space evolutionary model (SSEM) for the US-pop dataset. All errors are normalized by the values for the static model.}
\label{fig_Summary_DDMMPrediction_USPop}
\end{figure}
Next, we studied the prediction of future creation style distributions using the method described in Sec.~\ref{sec:PredictiveModelling}.
The experimental setups were the same as in Sec.~\ref{sec:EvaluationOfPredAccuracy}, and the referential year was set in the range $t_0\in [1979,1999]$.
The results of the DDMM parameter prediction using the SSEM in Fig.~\ref{fig_Summary_DDMMPrediction_USPop} show tendencies similar to those observed in Fig.~\ref{fig_Summary_DDMMPrediction_JPop} for the J-pop dataset.
Overall, the SSEM improved the prediction accuracy compared to the static model.
Additionally, the predictions by the GP were often less accurate than those of the static model, indicating the non-trivial nature of the prediction problem.

\begin{figure}
\centering
\includegraphics[width=1.\columnwidth]{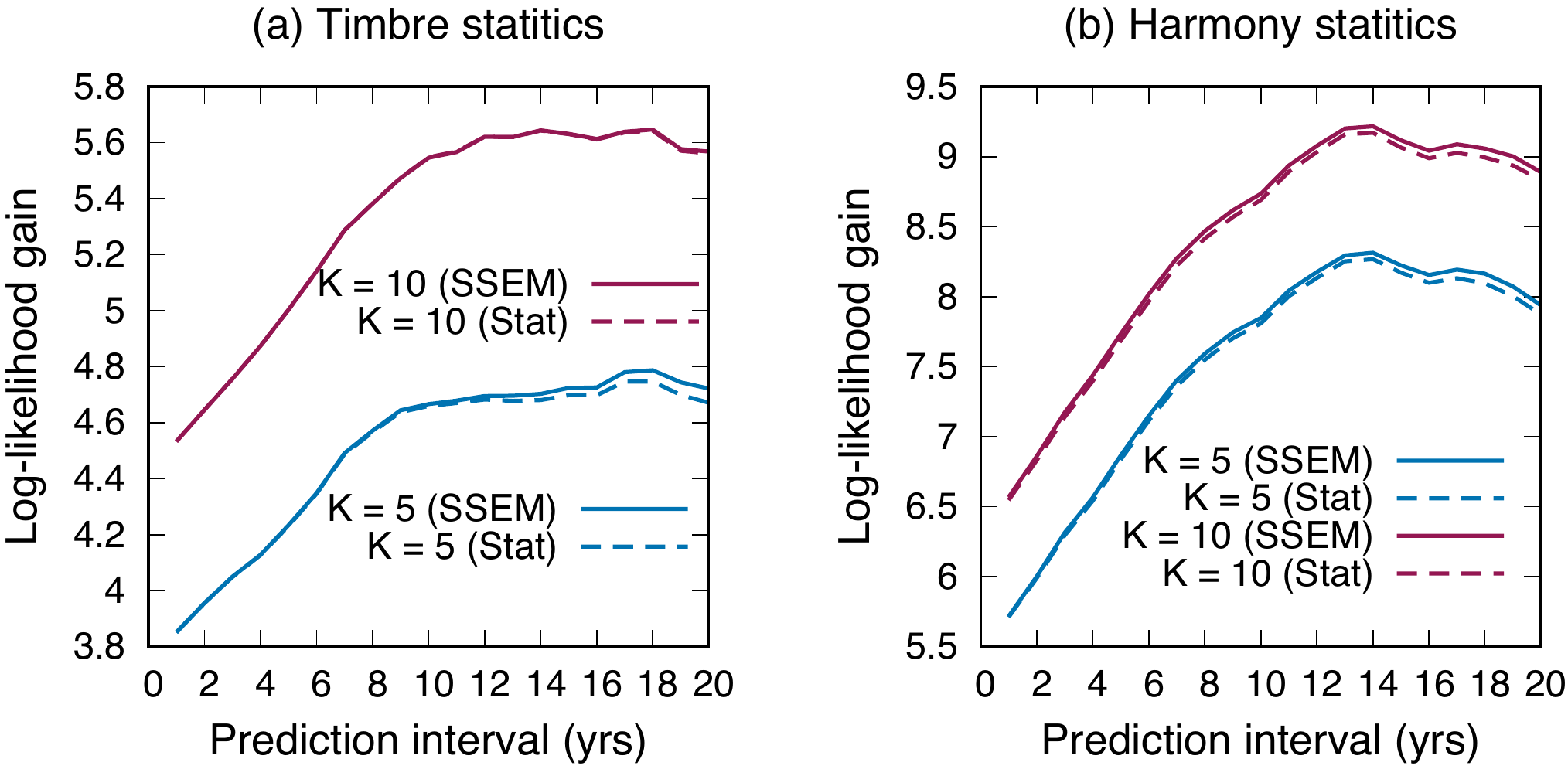}
\caption{The log-likelihood gains obtained by the SSEM and static model for the (a) timbre and (b) harmony statistics in the US-pop dataset.}
\label{fig_LogLikelihoodGain_USPop}
\end{figure}
\begin{figure}
\centering
\includegraphics[width=1.\columnwidth]{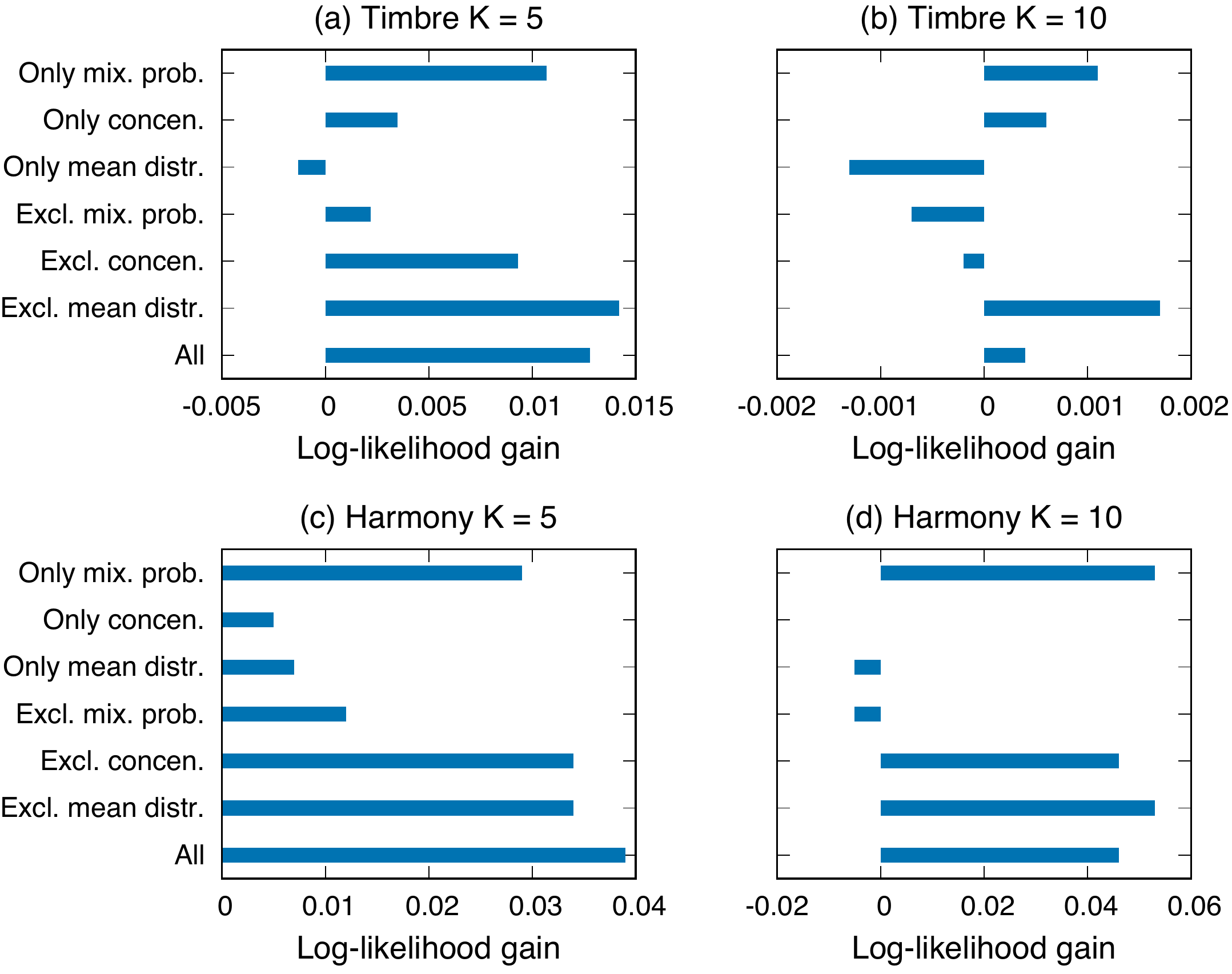}
\caption{The contributions of individual elements of the DDMM parameters to the log-likelihood gains for the US-pop dataset. (a) Timbre statistics ($K=5$), (b) timbre statistics ($K=10$), (c) harmony statistics ($K=5$), and (d) harmony statistics ($K=10$).}
\label{fig_ElementwiseContributionsToLogLikelihoodGain_USPop}
\end{figure}

The results of the likelihood evaluation (Fig.~\ref{fig_LogLikelihoodGain_USPop}) and the contributions of individual elements of the DDMM parameters to the likelihood gains (Fig.~\ref{fig_ElementwiseContributionsToLogLikelihoodGain_USPop}) indicate positive effects of the DDMM parameter predictions by the SSEM and the relevance of both inter- and intra-cluster dynamics, which are qualitatively the same conclusions as in the case of the J-pop dataset.
For the US-pop dataset, however, the log-likelihood gains obtained by the SSEM were relatively small, and using $K=10$ clusters always led to a higher prediction accuracy than using $K=5$ clusters.
The latter result can be ascribed to the large number of samples and more complex structure of the creation style distributions in this dataset.

\section{Discussion and conclusion}
\label{sec:Discussion}

In this study, we have investigated the evolution of music creation style distributions using statistical modelling methods.
By analyzing clusters found in several musical statistics extracted from the J-pop and US-pop datasets, we identified intra-cluster dynamics, such as the contractions of clusters' sizes and shift of clusters' centres, as well as prominent inter-cluster dynamics in the relative frequencies of clusters.
These different modes of cluster dynamics were incorporated into the dynamic Dirichlet mixture model (DDMM) in a unified manner.
We then used the DDMM to evaluate its predictive ability and found that both inter- and intra-cluster dynamics are relevant for predicting future creation style distributions.
In particular, these results highlight the importance of analyzing cluster variances for characterizing and predicting the evolution of creation style distributions, in addition to analyzing the frequencies and means of clusters, which have been the focus of most previous studies on quantitative cultural evolution analysis \cite{Griffiths2004,Michel2011,Mauch2015,Interiano2018}.
These conclusions were drawn for both datasets using different musical elements, implying that they possibly represent a cross-cultural universal nature of the evolution of musical culture.

Our results open new possibilities regarding the quantitative and theoretical understanding of cultural evolution.
First, as the results indicate that different clusters in the space of creation styles can have different modes of intra-cluster dynamics, future research should further investigate the cause for differentiating the characteristics of these dynamics.
Since a cluster in cultural data often corresponds to a stylistic genre of a culture, which often involves a specific community in a society, the type of the community's public preferences, such as preferences towards typicality or novelty \cite{Nakamura2019}, might provide a possible explanation.

Second, from Figs.~\ref{fig_JPop_Epc_Cluster}, \ref{fig_JPop_Met_Cluster}, \ref{fig_USPop_TLex_Cluster}, and \ref{fig_USPop_HLex_Cluster}, clusters of creation styles have a hierarchical structure and significantly overlap with each other.
In this situation, it is difficult to define the number of clusters, and therefore, the distinction between inter- and intra-cluster dynamics is essentially relative.
While using a fixed number of clusters for analysis, as in this study, often facilitates the interpretation of dynamics, another possibility is to apply the nonparametric Bayesian framework \cite{Antoniak1974,Rasmussen1999} to consider all possible numbers of clusters.
As we have confirmed the concurrent and transient cluster structures of creation styles, it would be interesting to compare the present method (dynamic analysis after clustering) with the methods for jointly conducting clustering and dynamic analysis \cite{Blei2006,Glynn2019}.
These topics will be investigated in future research.

Third, using the state-space evolutionary model (SSEM), predicting mean distributions is often more difficult than predicting the mixture probabilities and concentrations.
A possible reason for this is the high dimensionality of the data space of creation styles.
Additionally, from the analysis of the J-pop dataset (Fig.~\ref{fig_Prediction_JPop}), the evolutions of the component statistics often have musical relations and are not totally independent.
This suggests that the effective dimension of evolution is much lower than the feature dimension, similar to some observations in the biological evolution of phenotypes \cite{Frentz2015,Furusawa2018}.
Thus, an important problem that will be addressed in future work is to estimate the effective space of evolution from the data and construct an evolutionary model in the reduced space.

Finally, the present methodology, which analyzes the evolution of the statistics of a probabilistic generative model, enables not only the analysis and prediction of content features but also the generation of synthetic cultural artefacts in the styles of the past, present, and future, in line with the analysis-by-synthesis framework \cite{MacKay1951}.
This approach can therefore be applicable for realizing a ``creative'' system for music generation, rather than only a system imitating the style of existing artefacts, which is a major problem in automatic music generation systems based on machine learning \cite{Fernandez2013}.
Moreover, the proposed methodology is general and can potentially be applied to other domains of creative culture, such as lyrics \cite{Brand2019}, visual art \cite{Sigaki2018}, scientific papers \cite{Bergsma2012}, and culinary art \cite{Kinouchi2008}, where dynamical cluster structure and content-based selection play vital roles.

\section*{Data availability}

The music statistics data and source code used for the analysis, the data obtained by the analysis, and melody audio samples are available at\\
\url{https://drive.google.com/drive/folders/1n0nu89NmseIBIENMTFcB7PUq080MvkFb?usp=sharing} .

\section*{Acknowledgment}

This work is supported by JSPS KAKENHI Nos.\ 16J05486 and 22H03661, and ISHIZUE 2021 of Kyoto University Research Development Program.

\end{document}